\documentclass[11pt,fullpage]{article}

\long\def\ignore#1{}

\usepackage{graphicx}
\usepackage{latexsym}
\usepackage{color}
\usepackage{multirow}
\usepackage{amsmath}
\usepackage{amssymb}
\usepackage{amsthm}
\usepackage{amsfonts}
\usepackage{bm}

\usepackage[a4paper, total={6.5in, 9in}]{geometry}

\usepackage[utf8]{inputenc}

\usepackage[style=authoryear,dashed=false,backend=biber,citestyle=numeric,maxbibnames=99]{biblatex}
\renewbibmacro{in:}{}
\DeclareNameAlias{sortname}{last-first}
\DeclareFieldFormat{pages}{#1}
\DeclareFieldFormat{labelnumberwidth}{\mkbibbrackets{#1}}

\renewbibmacro*{volume+number+eid}{%
  \printfield{volume}%
  \setunit*{\addnbspace}
  \printfield{number}%
  \setunit{\addcomma\space}%
  \printfield{eid}}
\DeclareFieldFormat[article]{number}{\mkbibparens{#1}}

\defbibenvironment{bibliography}
  {\list
     {\printtext[labelnumberwidth]{%
      \printfield{labelprefix}%
      \printfield{labelnumber}}}
     {\setlength{\labelwidth}{\labelnumberwidth}%
      \setlength{\leftmargin}{\labelwidth}%
      \setlength{\labelsep}{\biblabelsep}%
      \addtolength{\leftmargin}{\labelsep}%
      \setlength{\itemsep}{\bibitemsep}%
      \setlength{\parsep}{\bibparsep}}%
      
      }
  {\endlist}
  {\item}
   
\usepackage{xcolor}
\usepackage{array}

\usepackage{mathtools}
\usepackage[ruled,vlined]{algorithm2e}
\usepackage{subcaption}
\usepackage{graphicx}
\usepackage{wrapfig}
\usepackage{float}
\usepackage[section]{placeins} 
\usepackage[font=scriptsize,labelfont=bf]{caption}
\usepackage{setspace}
\usepackage{amsfonts}
\usepackage{amsthm}
\usepackage{centernot}

\usepackage{amssymb}
\usepackage{graphicx}
\usepackage{latexsym}
\usepackage{multirow}

\DeclareMathOperator*{\argmax}{arg\,max}
\DeclareMathOperator*{\argmin}{arg\,min}

\newtheorem{assumption}{Assumption}[section]

\newtheorem{theorem}{Theorem}[section]
\graphicspath{ {images/} }
\begin{document}

\title{\bf High-dimensional regression with a count response}
\author{{\bf Or Zilberman} \\
Department of Statistics and Operations Research \\
Tel Aviv University \\ Israel \\
{\em orzilber@gmail.com} 
\and
{\bf Felix Abramovich} \\
Department of Statistics and Operations Research \\
Tel Aviv University \\
Israel \\
{\em felix@tauex.tau.ac.il} }

\date{}
\maketitle

\begin{abstract} We consider high-dimensional regression with a count response modeled by Poisson or negative binomial generalized linear model (GLM). We
propose a penalized maximum likelihood estimator with a properly chosen complexity penalty and establish its adaptive minimaxity across models of various sparsity. To make the procedure computationally feasible for high-dimensional data we consider its LASSO and SLOPE convex surrogates. Their performance is illustrated through simulated and real-data examples.
\end{abstract} 


\section{Introduction} \label{Intro}
Regression with a count response appears in a wide range of statistical applications, such as analysing
the number of trades in a given time interval, the number of queries in call center, the prolonged length of stay in hospital, the number of user's reviews on a given product, among others. It also an integral part of the analysis of contingency tables. A common model for a count response as a function of a set of predictive variables (features) is a Poisson regression. However, in practice, one often encounters the overdispersion phenomenon, where the variance of the response is larger than expected under the Poisson model. Overdispersion may be caused by different factors, such as the presence of other missing predictors, correlations between observations or outliers.
One common approach to modelling overdispersed count data is through the negative binomial (NB) distribution. Both Poisson and NB distributions belong to the natural exponential family allowing them to be considered within the general GLM framework developed in \cite{McCullagh:1989}. 

In the era of Big Data, a researcher deals with vast amounts of high-dimensional data. The number of features $d$ is large and may be even larger than the available sample size $n$ that raises a severe ``curse of dimensionality'' problem. Analysing the ``$d$ larger than $n$'' setup requires novel statistical approaches and techniques. Reducing the dimensionality of the feature space by selecting a sparse subset of ``significant'' features by feature/model selection
procedures becomes crucial. 

The use of various feature selection procedures for count data is not entirely new to practitioners. For example, \cite{Wang} and
\cite{10.1371/journal.pone.0209923} extended the well-known LASSO techniques to Poisson and NB regression for analysing respectively
the prolonged hospital length of stay after pediatric cardiac surgery and
micronuclei frequencies in genome studies.  \cite{Li:2022} considered a heterogeneous NB model  and used double NB regressions, accounting for both the dependent variable and the overdispersion parameter, by applying double LASSO for model selection techniques to both regressions. However, the underlying theoretical ground for high-dimensional count data has been much less developed. The goal of this paper is to bridge the gap.

The theory of model selection in Gaussian linear regression is well-developed (see \cite{var_sel_overview} and \cite{Giraud:2015} for reviews and references therein). \cite{minimax_in_glm} and \cite{High_dimension_classification} extended those results to high-dimensional GLM.
In this paper we adapt their general approach to present rigorous theoretical ground for model selection in Poisson and NB regression models.  In particular, we consider penalized maximum likelihood estimation with a complexity penalty on the model size. Utilizing the results of \cite{minimax_in_glm} for Poisson and NB regressions, we show that for a particular type of a penalty, the resulting estimator achieves the rate-optimal (in the minimax sense) Kullabck-Leibler risk simultaneously across models of various sparsity. 

Any model selection criterion based on a complexity penalty requires, however, a combinatorial search over all possible models, which is computationally infeasible for high-dimensional data. We therefore consider the LASSO (\cite{LASSO_PAPER}, \cite{LASSO_van_de_geer}) and its recent generalization SLOPE (\cite{FIRST_SLOPE}, \cite{SLOPE_PAPER}, \cite{High_dimension_classification}) penalties as convex surrogates for complexity penalties. Based on existing results for GLM, we show, that under certain additional (mild) conditions on the design, the SLOPE estimator with properly chosen tuning parameters is rate-optimal, while the LASSO estimator is sub-optimal. We also present the FISTA algorithm of \cite{FISTA} for numerical solving SLOPE for NB regression.  The performance of LASSO and SLOPE estimators is illustrated through a series of simulated and on a real-data example. 

The paper is organized as follows. Section \ref{sec:main} contains the main theoretical results of the paper. We present high-dimensional Poisson and negative binomial regression models for count data (Section \ref{subsec:model}), 
propose a penalized maximum likelihood estimator with a properly chosen complexity penalty and establish its minimaxity (Section \ref{subsec:complexity}). Its LASSO and SLOPE convex surrogates are investigated in
Section \ref{subsec:convex}. In Section \ref{sec:examples} we illustrate the performance of LASSO and SLOPE feature selection procedures through simulated and real-data examples. The FISTA algorithm for solving SLOPE in negative binomial regression is given in the Appendix.

\section{Theoretical results} \label{sec:main}

\subsection{Models for count data} \label{subsec:model}
Given a sample $(\mathbf{x}_i,Y_i),\;i=1,\ldots,n$, where $\mathbf{x}_i \in \mathbb{R}^d$ and $Y_i \in \mathbb{N} \cup \{0\}$ are counts, let $X \in \mathbb{R}^{n \times d}$ be the corresponding design matrix with rows $\mathbf{x}_i$ and $r=rank(X)$. Assume that any $r$ columns of $X$ are linearly independent.

A standard setup for count data is a Poisson log-linear regression model 
\begin{equation}\label{section_2:Poisson_model}
    Y_i \sim Pois(\lambda_i), \;\;\; \ln\lambda_i = \mathbf{\beta}^t \mathbf{x}_i
\end{equation}
with the  log-likelihood  
\begin{equation} \label{eq:loglike_Poisson}
    l(\mathbf{\beta}) = \sum\limits_{i = 1}^n\left(\tilde{\beta}^t \mathbf{x_i} Y_i - e^{\tilde{\beta}^t \mathbf{x}_i}\right) .
\end{equation}

The Poisson regression model (\ref{section_2:Poisson_model}) implies
$EY_i=Var(Y_i)=\lambda_i$. However, as we have mentioned in the introduction, analysing count data one often encounters the overdispersion phenomenon, where $Var(Y_i)>EY_i$. One common approach to modelling overdispersed count data is through the NB regression model:
\begin{equation}\label{section_2:NB_model}
Y_i  \sim NB\left(\alpha,\frac{\alpha}{\alpha+\lambda_i}\right),\;\; \;\ln \lambda_i  = \mathbf{\beta}^t \mathbf{x}_i,
\end{equation}
where $\alpha$ is a nuisance parameter. 
The corresponding log-likelihood is
\begin{equation} \label{eq:loglike_NB}
l(\mathbf{\beta}) = \sum\limits_{i = 1}^n\left(Y_i\ln\left(\frac{e^{\mathbf{\beta}^t \mathbf{x}_i}}{e^{\mathbf{\beta}^t  \mathbf{x}_i}+\alpha}\right) + \alpha \ln\left(\frac{\alpha}{e^{\mathbf{\beta}^t \mathbf{x}_i}+\alpha}\right)\right) .
\end{equation}

For the NB regression model (\ref{section_2:NB_model}), $EY_i = \lambda_i$, while $Var(Y_i) = \lambda_i + \frac{\lambda_i^2}{\alpha} > \lambda_i$.
The NB distribution can be viewed as a $Gamma(\alpha,\alpha)$ mixture of Poisson distributions and approaches the Poisson distribution as $\alpha$ increases. 

Both Poisson and NB distributions belong to the natural exponential family that allows one to use the general GLM methodology and techniques.
Thus, although the maximum likelihood estimator (MLE) $\hat{\beta}$ of $\beta$ in (\ref{eq:loglike_Poisson}) and (\ref{eq:loglike_NB}) is not available in closed form, it can nevertheless
be obtained numerically by the iteratively reweighted least squares algorithm 
(see \cite{McCullagh:1989}, Section 2.5).

Assume the following boundedness assumption: 
\begin{assumption} \label{ass:A}
Consider the models (\ref{section_2:Poisson_model}) or (\ref{section_2:NB_model}). Assume that there exists $C_0>0$ such that $|\mathbf{\beta}^t \mathbf{x}_i | \leq C_0 $ for all $i=1,\ldots,n$.
\end{assumption}
Assumption \ref{ass:A} prevents the variances $Var(Y_i)$ to be infinitely large or arbitrarily close to zero, where MLE may fail.

\subsection{Feature selection by penalized MLE with complexity penalties} 
\label{subsec:complexity}
Recall that we are especially interested in high-dimensional setups, where $d$ is large w.r.t. $n$ and feature (model) selection is essential. We assume that the true underlying regression model is sparse with $||\mathbf{\beta}||_0=d_0 < d$, 
where the $l_0$ (quasi)-norm of $\mathbf{\beta}$ is the number of its nonzero entries, and consider 
penalized maximum likelihood model selection approach.

For a given model $M \subseteq \{1,\ldots,d\}$ let $\mathcal{B}_M = \{\mathbf{\beta} \in \mathbb{R}^d : \beta_j = 0\;{\rm if}\; j \notin M \}$ and consider the corresponding MLE $\hat{\beta}_M$ of $\mathbf{\beta}$:  
\begin{equation}
    \hat{\beta}_{M} = \argmax_{\tilde{\beta} \in \mathcal{B}_M} l(\tilde{\beta}), 
\end{equation}
where the log-likelihood $l(\tilde{\beta})$ is given in (\ref{eq:loglike_Poisson}) or (\ref{eq:loglike_NB}).

Let $\mathfrak{M}$ be the set of all $2^d$ possible models and select the model $\widehat{M}$ by the penalized MLE criterion of the form
\begin{equation} \label{eq:Global_PMLE}
    \widehat{M} = \argmin_{M \in \mathfrak{M}}\{-l(\hat{\mathbf{\beta}}_M) + Pen(|M|)\}
\end{equation}
with the complexity penalty $Pen(\cdot)$ on the model size $|M|=||\hat{\mathbf{\beta}}_M||_0$. The resulting estimator of $\mathbf{\beta}$ is $\hat{\mathbf{\beta}}_{\widehat M}$. In fact, we can restrict $\mathfrak{M}$ in (\ref{eq:Global_PMLE})  to models with sizes at most $r$ since for any $\mathbf{\beta}$ with $||\mathbf{\beta}||_0 > r$, there necessarily exists another $\mathbf{\beta}'$ with $||\mathbf{\beta}'||_0 \leq r$ such that $X\mathbf{\beta}=X\mathbf{\beta}'$. 

Such an approach is well-known and depends on the chosen complexity penalty function $Pen(\cdot)$. Thus, the widely-used AIC \cite{AIC_paper}, BIC \cite{dim_of_model} and RIC \cite{RIC_paper} criteria correspond to {\em linear} penalties of the form $Pen(|M|) = C|M|$ with $C=1$ for AIC, $C = (\ln n)/2$ for BIC and $C=\ln d$ for RIC respectively.  
\cite{minimax_in_glm} considered a {\em nonlinear} complexity penalty of the form $Pen(|M|) = C |M|\ln(\frac{de}{|M|})$ for a general GLM setup. Such a penalty behaves similar to a conservative RIC for sparse models with $|M| \ll d$ and to a very liberal AIC for dense models with $|M| \sim d$.

Let $f_{\lambda_1}(y)$ and $f_{\lambda_2}(y)$ be two Poisson or NB probability distributions with parameters $\lambda_1$ and $\lambda_2$ correspondingly. A Kullback-Leibler divergence $KL(\lambda_1,\lambda_2)=\mathbb{E}_{\lambda_1}\ln\left( \frac{f_{\lambda_1}(Y)}{f_{\lambda_2}(Y)}\right)$, and straightforward calculus yields
\begin{equation} \label{eq:KLPoisson}
KL(\lambda_1,\lambda_2)=\lambda_1 \ln\frac{\lambda_1}{\lambda_2}-\lambda_1+\lambda_2
\end{equation}
for Poisson and
\begin{equation} \label{eq:KLNB}
KL(\lambda_1,\lambda_2)=\lambda_1\left(\ln \frac{\lambda_1}{\lambda_1+\alpha}-\ln \frac{\lambda_2}{\lambda_2+\alpha}\right)+\alpha \ln \frac{\lambda_1+\alpha}{\lambda_2+\alpha}
\end{equation}
for NB distributions.

We measure the goodness-of-fit of $\hat{\mathbf{\beta}}$
by the expected  Kullback-Leibler divergence (Kullback-Leibler risk) $\mathbb{E}KL(X\mathbf{\beta},X\hat{\mathbf{\beta}}_{\widehat{M}})$ 
 between the true data distribution and its empirical distribution generated by $\hat{\lambda}_{\widehat M}=e^{X\hat{\mathbf{\beta}}_{\widehat M}}$. From (\ref{eq:KLPoisson}) and (\ref{eq:KLNB}) we have
\begin{equation}\label{eq:Poisson reg KL}
    KL(X\beta,X\hat{\beta}_{\widehat M}) = \sum\limits_{i=1}^n\left(\exp{(\mathbf{\beta}^t \mathbf{x}_i)} (\mathbf{\beta}-\hat{\mathbf \beta}_{\widehat M})^t \mathbf{x}_i - (\exp{(\hat{\mathbf \beta}^t \mathbf{x}_i)} - \exp{(\hat{\mathbf{\beta}}_{\widehat M}^t \mathbf{x}_i)})\right)
\end{equation}
and
\begin{equation}\label{eq:NB reg KL}
\begin{split}
    KL \big(X\mathbf{\beta},X \mathbf{\hat{\beta}}_{\widehat M}\big)  = & 
    \sum\limits_{i=1}^n \left(\exp(\mathbf{\beta}^t \mathbf{x}_i) \bigg(\ln\frac{\exp(\mathbf{\beta}^t \mathbf{x}_i)}{\exp(\mathbf{\beta}^t \mathbf{x}_i)+\alpha} - \ln\frac{\exp(\hat{\mathbf{\beta}}_{\widehat M}^t \mathbf{x}_i)}{\exp(\mathbf{\hat{\beta}}_{\widehat M}^t \mathbf{x}_i)+\alpha} \bigg) \right. \\ & + \left. \alpha \ln\frac{\exp(\mathbf{\hat{\beta}}_{\widehat M}^t \mathbf{x}_i)+\alpha}{\exp(\mathbf{\beta^t} \mathbf{x}_i)+\alpha} \right)
\end{split}
\end{equation}
for the Poisson (\ref{section_2:Poisson_model}) and NB (\ref{section_2:NB_model}) regression models respectively. 

The following theorem is a direct consequence of the results of  \cite{minimax_in_glm} for a general GLM setup adapted to Poisson and NB log-linear regression models:
\begin{theorem} \label{th:KLbound}
Consider a $d_0$-sparse Poisson (\ref{section_2:Poisson_model}) or NB (\ref{section_2:NB_model}) regression model, where $||\mathbf{\beta}||_0 \leq d_0$, $1 \leq d_0 \leq r$. Apply the model selection
procedure (\ref{eq:Global_PMLE}) with the complexity penalty
\begin{equation} \label{eq:penalty}
Pen(k)=C k\ln\left(\frac{de}{k}\right),\;\;\;k=1,\ldots,r-1\;\;\;{\rm and}\;\;\;
Pen(r)=C r,
\end{equation}
where the constant $C>0$ is given in \cite{minimax_in_glm}. 

Then, under Assumption \ref{ass:A},
\begin{equation} \label{eq:KL_upper}
\sup_{\mathbf{\beta}: ||\mathbf{\beta}||_0 \leq d_0}
\mathbb{E}KL(X\mathbf{\beta},X\hat{\mathbf{\beta}}_{\widehat{M}})=O\left( \min\left(d_0 \ln\left( \frac{de}{d_0}\right),r\right)\right)
\end{equation}
Moreover, the upper bound (\ref{eq:KL_upper}), up to probably a different constant, is also the minimax bound for Kullback-Leibler risk among the set of $d_0$-sparse models $\mathcal{B}(d_0)=\{\mathbf{\beta} \in \mathbb{R}^d: ||\mathbf{\beta}||_0 \leq d_0\}$.
\end{theorem}
Theorem \ref{th:KLbound} shows that the penalized MLE with the complexity penalty (\ref{eq:penalty}) 
is simultaneously minimax for all $1 \leq d_0 \leq r$.

From \cite{minimax_in_glm} it also follows that the RIC-type model selection procedure with a linear complexity penalty of the form $C (\ln d) |M|$ is sub-optimal with the Kullback-Leibler risk of the order $O\left(d_0 \ln d\right)$.

\subsection{Convex relaxation} \label{subsec:convex}
Despite the strong theoretical results from the previous section, the practical implementation of model selection procedures based on the penalized MLE with a complexity penalty faces a serious computational drawback. Solving (\ref{eq:Global_PMLE}) requires a combinatorial search over all $2^d$ possible models that makes it computationally infeasible for high-dimensional data. One should apply then a convex relaxation techniques to replace the original complexity penalty in (\ref{eq:Global_PMLE}) by a related convex surrogate.
The celebrated LASSO \cite{LASSO_PAPER} replaces the non-convex $l_0$-(quasi)-norm in the linear complexity penalty by the convex $l_1$-norm:
\begin{equation} \label{eq:LASSO}
\hat{\mathbf{\beta}}_L=\arg \min_{\tilde{\mathbf{\beta}}} \left\{-l(\tilde{\mathbf{\beta}})+\gamma ||\tilde{\mathbf{\beta}}||_1\right\},
\end{equation}
where $\gamma$ is a tuning parameter. Solving (\ref{eq:LASSO}) implies a sparse solution, where part of the entries of $\hat{\mathbf{\beta}}_L$ are zeroes.

Assume that the columns of the design matrix $X$ are normalized to have unit Euclidean norms.
Convex relaxation requires certain extra conditions on the restricted minimal eigenvalues of the design matrix $X$  (see \cite{SLOPE_PAPER} for discussion). 
\cite{LASSO_van_de_geer} showed that under those conditions, the LASSO with a tuning parameter $\gamma \sim \sqrt{2 \ln d}$, similar to RIC, achieves a sub-optimal (up to a different log-factor) Kullback-Leibler risk of the order $O(d_{0}\ln d)$ in a general GLM setup. 
Similar results were obtained in \cite{SPRWL} for square root LASSO estimator in sparse Poisson regression.

A recently proposed generalization of LASSO is SLOPE  that utilizes a  \textit{sorted} $l_1$-norm penalty:
\begin{equation} \label{eq:slope}
   \hat{\mathbf{\beta}}_S=\arg \min_{\tilde{\mathbf{\beta}}} \left\{-l(\tilde{\mathbf{\beta}})+\sum_{j=1}^d \gamma_j |\tilde{\beta}|_{(j)}\right\}, 
\end{equation}
where $|\tilde{\beta}|_{(1)} \geq \ldots \geq |\tilde{\beta}|_{(d)}$ are the ordered absolute values of $\tilde{\beta}_j$'s and $\gamma_1 \geq \ldots \geq \gamma_d > 0$ are the tuning parameters. Evidently, LASSO is a particular case of SLOPE for equal $\gamma_j$'s. SLOPE was originated in 
\cite{FIRST_SLOPE} for Gaussian data and extended for GLM in \cite{High_dimension_classification}.

Following \cite{SLOPE_PAPER}, assume the {\em weighted restricted eigenvalue} (WRE) condition for the SLOPE estimator (\ref{eq:slope})~:
\begin{assumption}{($WRE(d_0,c_0)$ condition)}
Consider the $d_0$-sparse Poisson (\ref{section_2:Poisson_model}) or NB (\ref{section_2:NB_model}) regression model  with 
$||\mathbf{\beta}||_0 \leq d_0$, where the columns of the design matrix $X$ are normalized to have unit Euclidean norms.
Consider the cone of vectors ${\cal S}(d_0,c_0)=\{\mathbf{u} \in \mathbb{R}^d: \sum_{j=1}^d \sqrt{\ln(2d/j)} |u|_{(j)} 
\leq (1+c_0) ||\mathbf{u}||_2~ \sqrt{\sum_{j=1}^{d_0} \ln(2d/j)} \}$ for $c_0>0$ and assume that
$$
\kappa(d_0)=\inf_{\mathbf{u} \in {\cal S}(d_0,c_0) \backslash \{\mathbf{0}\}}
\frac{||X\mathbf{u}||^2}{||\mathbf{u}||^2} >0
$$
\end{assumption}

The relations between the WRE condition and the restricted eigenvalue condition required for LASSO are discussed in \cite{SLOPE_PAPER}.

Adapting the general GLM results of \cite{High_dimension_classification}, Theorem 8 to Poisson and NB regression models implies the following result: 
\begin{theorem} \label{th:slope}
Consider a sparse $d_0$-sparse Poisson (\ref{section_2:Poisson_model}) or NB (\ref{section_2:NB_model}) regression model, where $||\mathbf{\beta}||_0 \leq d_0$, the columns of the design matrix $X$ are normalized to have unit Euclidean norms and, in addition, $X$ satisfies the $WRE(d_0,c_0)$ condition for 
some $c_0>1$. 

Consider the SLOPE estimator (\ref{eq:slope}) with the tuning parameters 
\begin{equation} \label{eq:gamma}
\gamma_j= \tilde{C} \sqrt{\ln(2d/j)},\;\;\;j=1,\ldots,d,
\end{equation}
where the constant $\tilde{C}>0$ is given in \cite{High_dimension_classification}.

Then, under Assumption \ref{ass:A},
\begin{equation}  \label{eq:klslope}
\sup_{\mathbf{\beta}: ||\mathbf{\beta}||_0 \leq d_0} 
\mathbb{E}KL(X\mathbf{\beta},X\hat{\mathbf{\beta}}_{\widehat{M}}) \leq C_1 \frac{d_0}{\kappa(d_0)}
\ln\left( \frac{de}{d_0}\right) 
\end{equation}
for some $C>0$.
\end{theorem}
Thus, the SLOPE estimator is both computationally feasible and rate-optimal (in the minimax sense) for all but very dense models where 
$d_0\ln(\frac{de}{d_0})>r$. In fact,  the SLOPE penalty with $\gamma_j$'s in (\ref{eq:gamma}) can be viewed as a convex surrogate for the nonlinear complexity penalty (\ref{eq:penalty}). 

\section{Examples} \label{sec:examples}

\subsection{Simulation study} \label{subsec:simulation}
To test the performance of LASSO and SLOPE procedures we conducted a simulation study based on that of \cite{Wang}.
We also compared them with a forward selection procedure -- a stepwise approximation of a linear complexity penalty AIC/RIC.

\subsubsection{Data generation}
Following \cite{Wang}, feature vectors $\mathbf{x} \in \mathbb{R}^d$ were independently drawn from a multivariate normal distribution $N_d(0,\Sigma)$ with $\Sigma_{ij}=\rho^{|i-j|},\;i,j= 1,\dotsc,d$, and were further normalized to have unit Euclidean norms. A subset of $d_0 \leq d$ active features was randomly chosen from the total $d$-dimensional feature set, where $d_0$ was defined in terms of the proportion of active features $\epsilon=d_0/\min(d,n_{train})$. We kept $d_0 \leq n_{train}$ to avoid non-identifiable models. The corresponding $d_0$ nonzero coefficients were randomly set to be $\pm 0.5$ or $\pm 0.6$. Due to the autoregressive correlations between predictors, random choice of $d_0$ predictors in each simulation introduced additional variability in generating the dependent variable allowing us to consider different correlation setups. Finally, for a given feature vector $\mathbf{x}$ and a coefficients vector $\mathbf{\beta}$ we generated Poisson response $Y \sim Pois(\lambda(\mathbf{x}))$ with $\mathbf{\lambda}(\mathbf{x})=e^{\mathbf{\beta}^t \mathbf{x}}$.

In the study we tried all combinations of $\rho = \{0,\;0.5,\;0.8\}$, $d=\{20,\;100,\;200,\;500,\;1000\}$ and $\epsilon=\{0.05,\;0.1,\;0.15,\;0.2,\;0.25,\;0.3,\;0.35,\;0.5,\;0.7,\;0.9\}$ to cover various sparse and dense scenarios. For each combination of $d,\; \epsilon$ and $\rho$ we simulated 100 multinormal vectors $\mathbf{x} \sim N(0,\Sigma_d)$ with $\Sigma_{ij} = \rho^{|i-j|}$ and $d_0$-sparse vectors $\mathbf{\beta}$ as described above. Following \cite{Wang}, for each $\mathbf{X}$ and $\mathbf{\beta}$ we generated then 300 random samples  from Poisson log-linear regression model with $\mathbf{\lambda}=e^{\mathbf{X}\mathbf{\beta}}$ and randomly split them to train ($n_{train}=200$) and test ($n_{test}=100$) sets.

\subsubsection{Numerical methods}
Both LASSO and SLOPE can be solved numerically by various convex optimization algorithms. 
More specifically, for Poisson and NB setups,
\cite{Wang} combined the iteratively reweighted least squares algorithm of \cite{McCullagh:1989} (Section 2.5) for fitting GLMs and the coordinate descent algorithm for solving LASSO. A similar approach is implemented in the R-package \textit{glmnet} of \cite{glmnet_paper} used in our study with $\gamma=C_L \sqrt{2 \ln d}$.

Unlike LASSO, SLOPE has emerged quite recently. \cite{FIRST_SLOPE} applied  the Fast Iterative Shrinkage-Thresholding Algorithm (FISTA) of \cite{FISTA} to solve SLOPE for Gaussian regression. \cite{Larsson:2022} extended it to general GLM, including Poisson regression as a particular case (see also Appendix).
We utilized their R-package \textit{SLOPE} \cite{SLOPE_package} to run  SLOPE with $\gamma_j=C_S \sqrt{\ln(2d/j)}$ (see (\ref{eq:gamma})).

The tuning parameters $C_L$ for LASSO and $C_S$ for SLOPE were chosen by a 5-fold cross-validation. 

Forward selection with the linear penalty $Pen(|M|)=C_F |M|$ was performed by \textit{stepAIC} function from the R-package \textit{MASS} \cite{MASS}, where the tuning constant $C_F$ was also chosen by cross-validation.

\subsubsection{Results}
\begin{figure}[h]
\centering
\includegraphics[clip,width=0.75\linewidth]{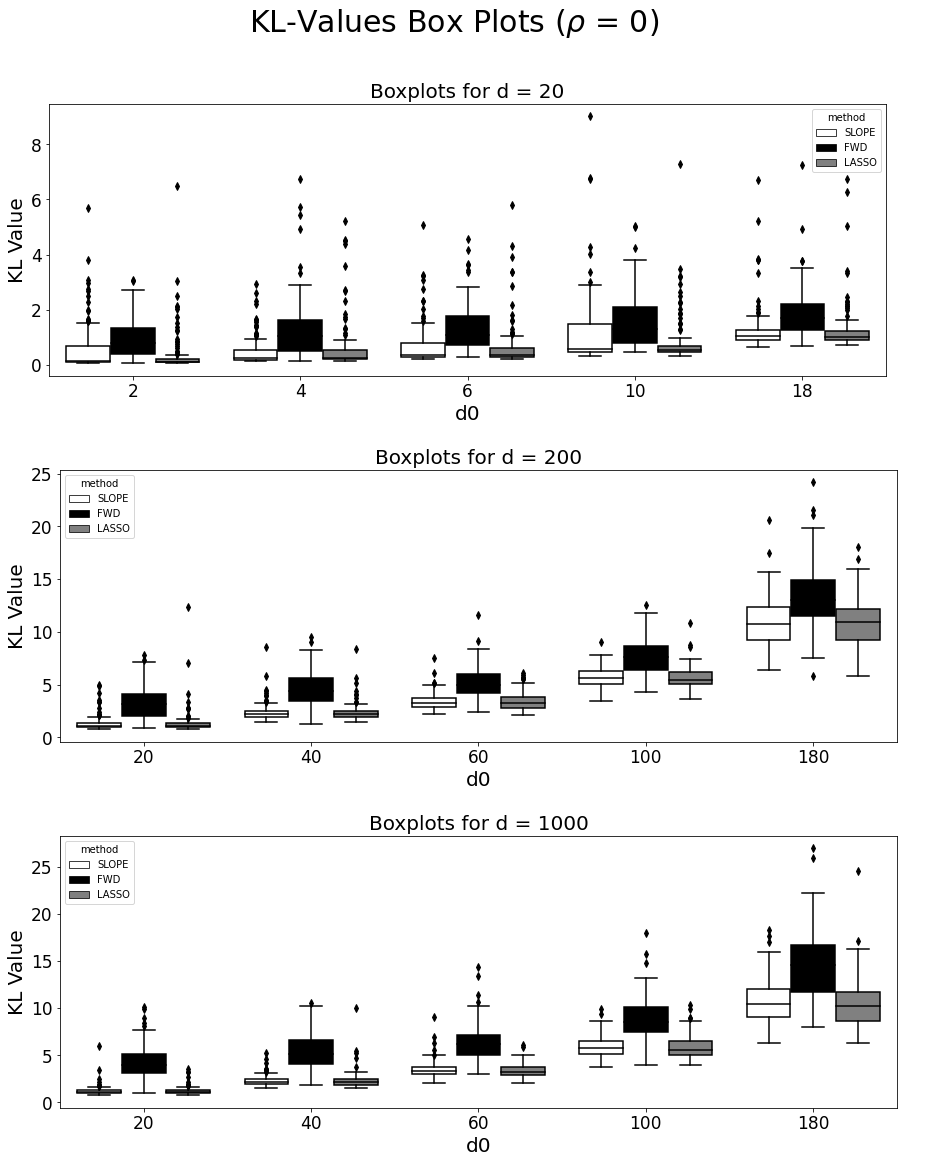}
\caption{Boxplots of KL-divergence for various $d$ and $d_0$, $\rho = 0$}
\label{boxplots_kl_0}
\end{figure}

\begin{figure}[h]
\includegraphics[clip,width=1\linewidth]{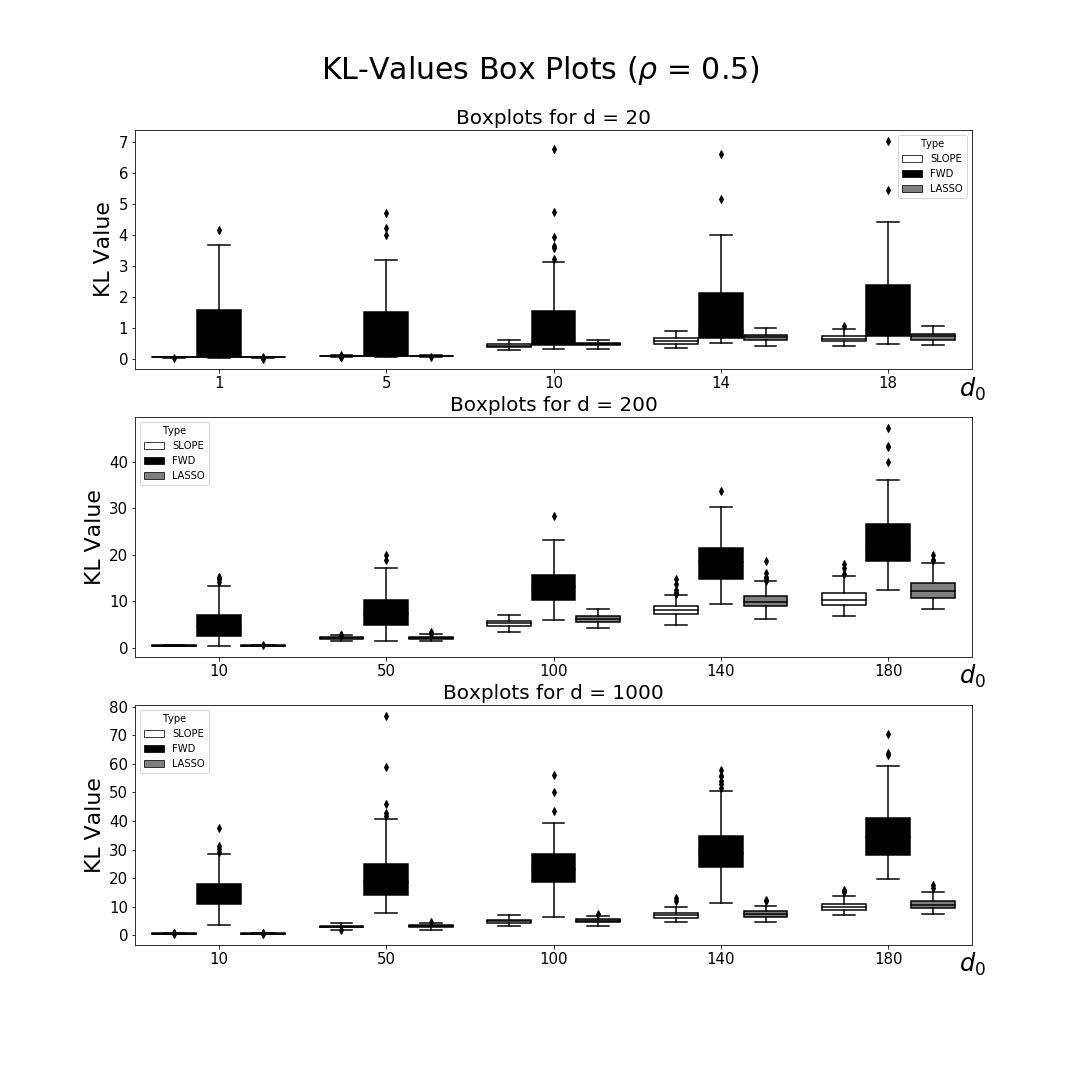}
\caption{Boxplots of KL-divergence for various $d$ and $d_0$, $\rho = 0.5$}
\label{boxplots_kl_0.5}
\end{figure}

\begin{figure}[h]
\centering
\includegraphics[clip,width=1\linewidth]{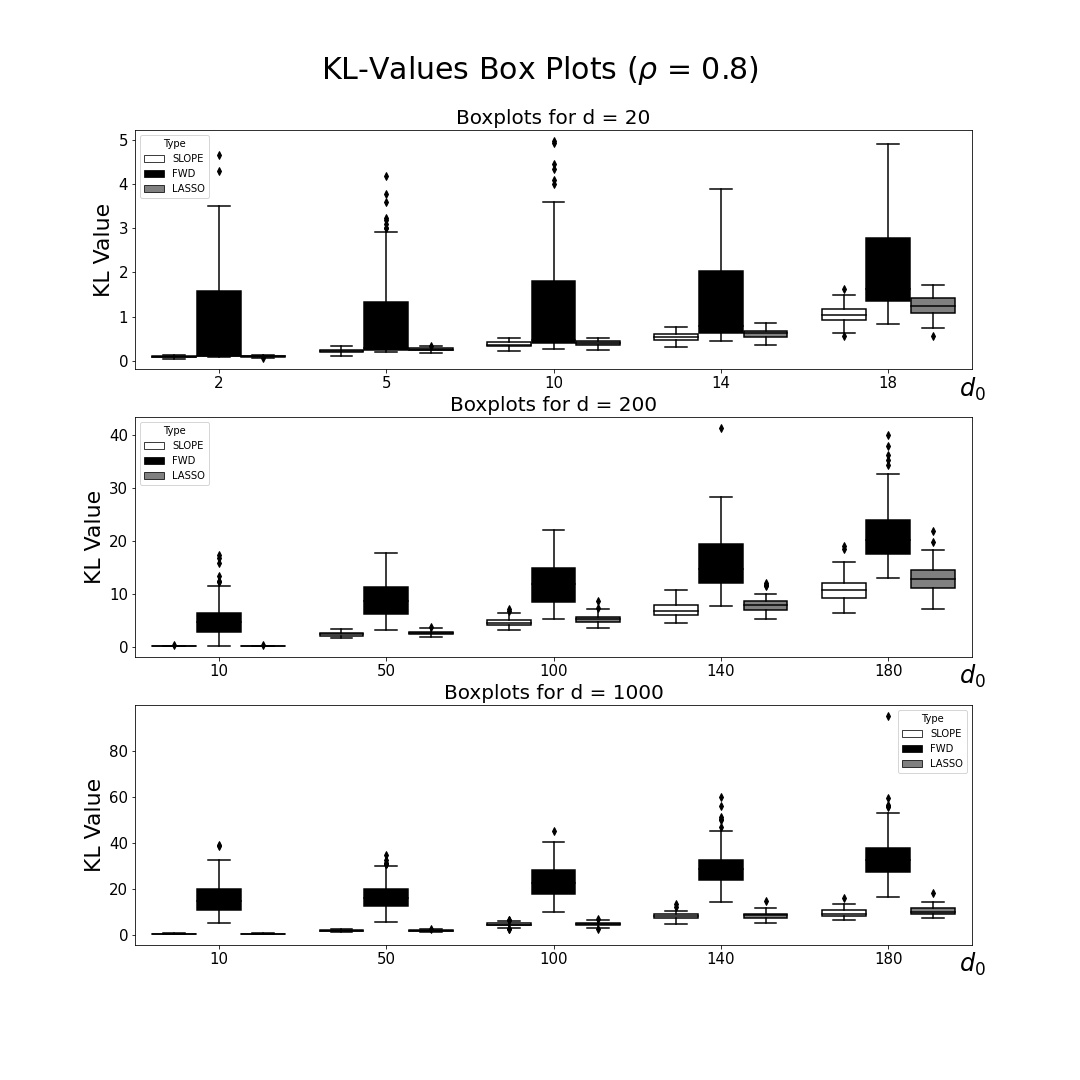}
\caption{Boxplots of KL-divergence for various $d$ and $d_0$, $\rho = 0.8$}
\label{boxplots_kl_0.8}
\end{figure}

For each simulation set, we calculated KL-divergences $KL({\bf Y}_{test},X_{test}\hat{\beta})$, where from  (\ref{eq:Poisson reg KL}), 
\begin{equation} \nonumber
\begin{split} 
KL({\bf Y}_{test},X_{test}\hat{\beta}) & =\sum_{i=1}^{n_{test}} \left(Y_{{test},i} \ln\left(\frac{Y_{{test},i}}{\exp(\hat{\beta}^t {\bf x}_{{test},i})}\right)-Y_{{test},i}+\exp(\hat{\beta}^t {\bf x}_{{test},i})\right) \\
& =\sum_{i=1}^{n_{test}} Y_{{test},i}\left(\ln\frac{Y_{{test},i}}{\exp(\hat{\beta}^t {\bf x}_{{test},i})}\right)
\end{split}  
\end{equation}
(the last equality assumes that there is an intercept in the selected model), which is actually half the deviance $D(X_{test}\hat{\beta})$ of the selected model on the test set. Figures \ref{boxplots_kl_0}--\ref{boxplots_kl_0.8}
present the boxplots for the resulting 
KL-divergences $KL({\bf Y}_{test},X_{test}\hat{\beta})$
obtained by the three methods for different $d, d_0$ and $\rho$. The figures show that the KL-divergence increases with both $d_0$ and $d$ across all methods. SLOPE and LASSO uniformly strongly outperform forward selection procedure with SLOPE typically resulting in smaller KL-divergence values than LASSO. Moreover, forward selection exhibits the largest dispersion in KL-divergence values indicating  its sensitivity to random samples, especially for correlated predictors. Superiority of SLOPE over LASSO in correlated settings grows with $\epsilon=\frac{d_0}{d}$, which is consistent with theoretical results from Section \ref{subsec:convex} for dense models.   

\FloatBarrier

\begin{figure}[h]
\centering
\includegraphics[clip,width=0.75\linewidth]{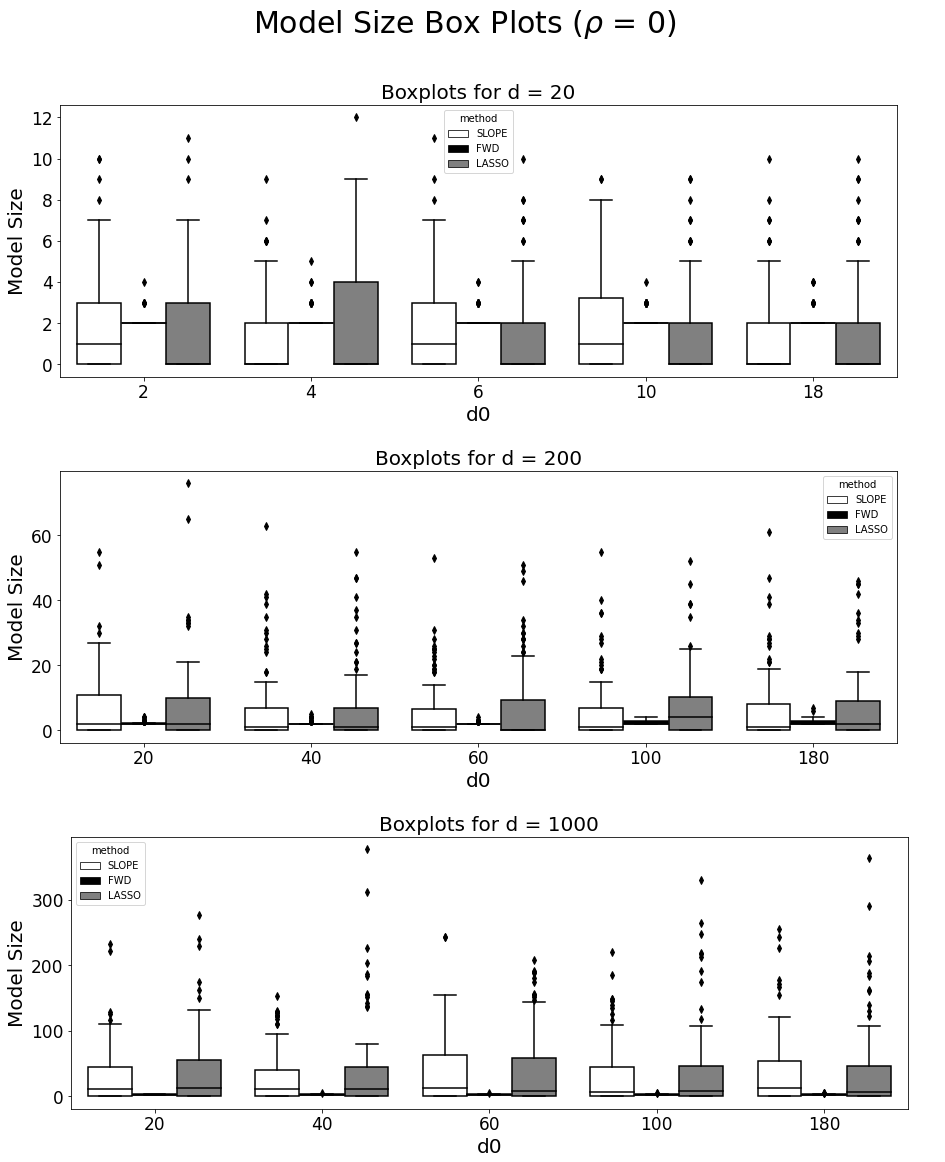}
\caption{Boxplots of selected  models' sizes for various $d$ and $d_0$ values, $\rho = 0$.}
\label{boxplots_MS_selected_0}
\end{figure}

\begin{figure}[h]
\centering
\includegraphics[clip,width=1\linewidth]{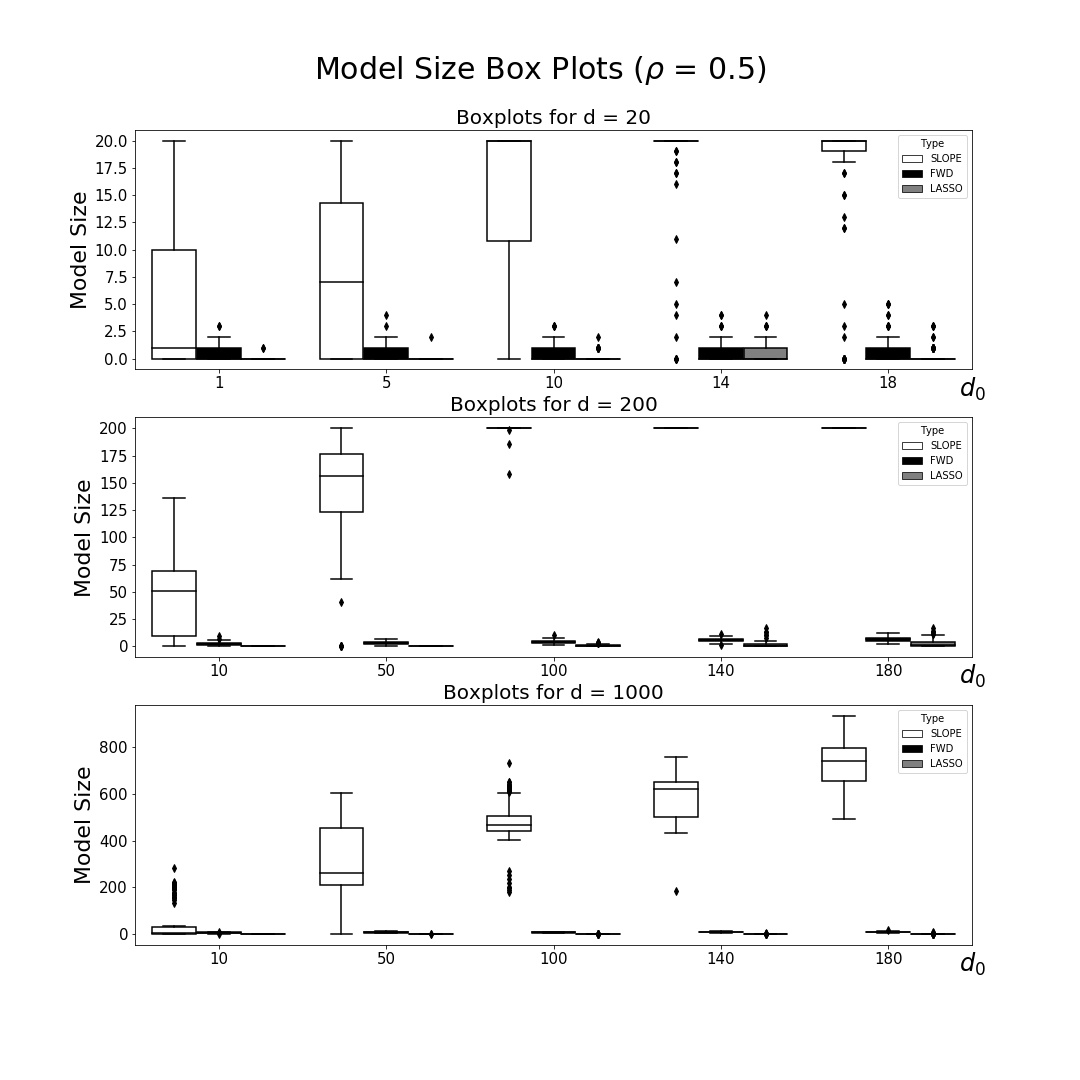}
\caption{Boxplots of selected models' sizes for various $d$ and $d_0$ values, $\rho = 0.5$.}
\label{boxplots_MS_selected_0.5}
\end{figure}

\begin{figure}[h]
\centering
\includegraphics[clip,width=1\linewidth]{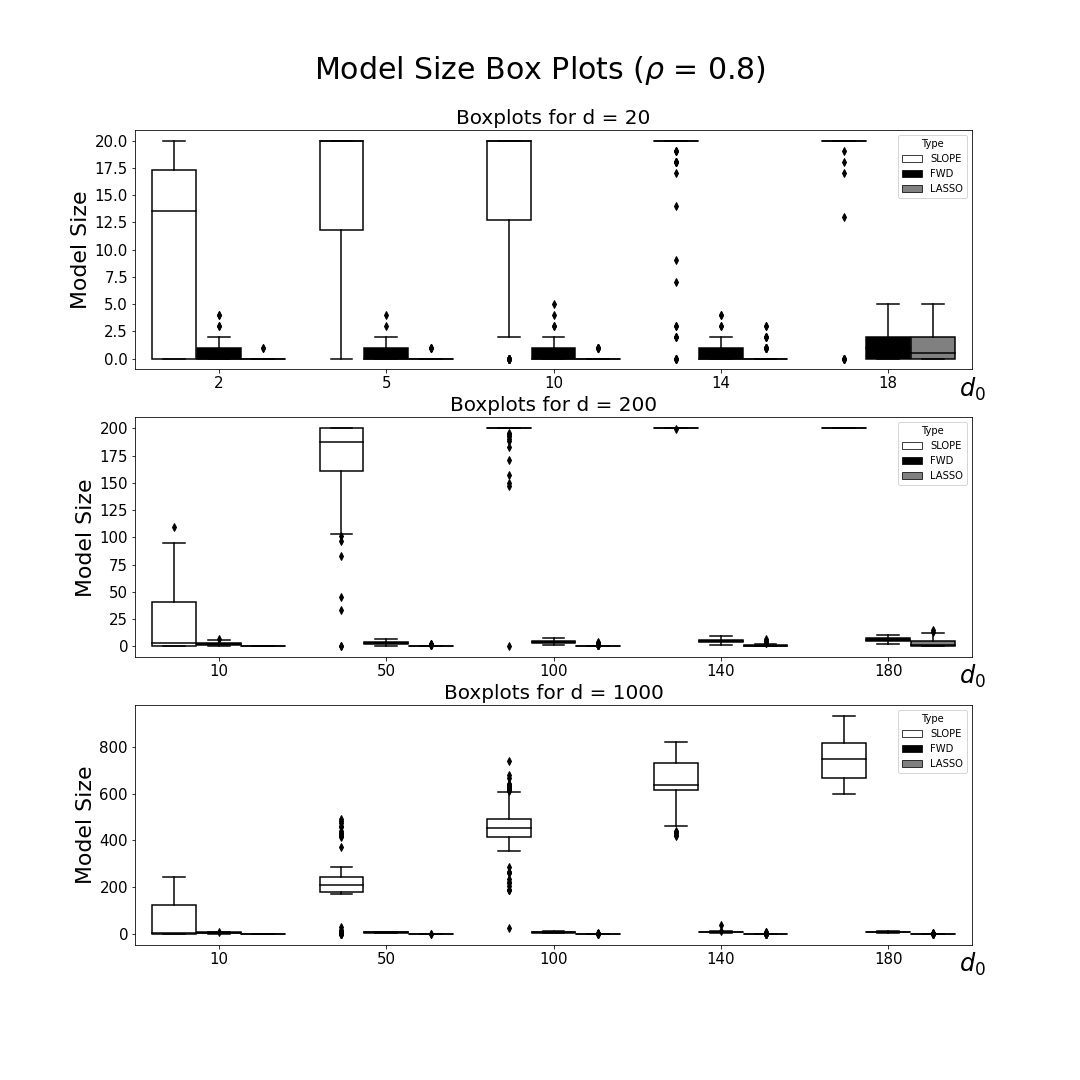}
\caption{Boxplots of selected models' sizes for various $d$ and $d_0$ values, $\rho = 0.8$.}
\label{boxplots_MS_selected_0.8}
\end{figure}

In addition to the KL-divergence, we compared the resulting model sizes for the three methods.  SLOPE and LASSO yielded similar ranges of model sizes in an uncorrelated setup (see Figure \ref{boxplots_MS_selected_0}). However, SLOPE consistently selected significantly larger models for correlated predictors (see Figures \ref{boxplots_MS_selected_0.5} and \ref{boxplots_MS_selected_0.8}), with much more dispersed model sizes. In fact, the phenomenon of SLOPE selecting large models in the  presence of correlated predictors  is  well-known. Thus, \cite{Figueiredo} showed that in the presence of a group (cluster) of highly correlated features, SLOPE tends to include all of them in the model. In contrast, LASSO usually selects only a small subset of features from the cluster (see \cite{Figueiredo} and \cite{Zou05regularizationand}) that sometimes might even cause over-sparsity. 
Forward selection provides the sparsest models for uncorrelated predictors, while LASSO provides the sparsest models otherwise.

\FloatBarrier
   
Summarizing, we can conclude that convex relaxation methods (LASSO and SLOPE) strongly outperform forward selection for prediction, with SLOPE providing the best results for correlated setups and for dense cases.
On the other hand, SLOPE tends to select larger models.

\subsection{Real data example}\label{subsec:real example}
\begin{figure}[h]
\includegraphics[clip,width=1\linewidth]{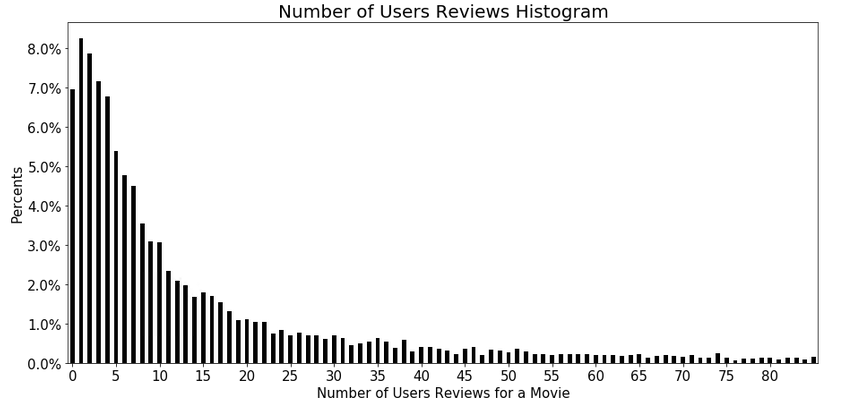}
\caption{Histograms of user's reviews (up to the 85th quantile).}
\label{fig:User_rev_hist}
\end{figure}

We also examined the performance of LASSO, SLOPE  and forward selection by analyzing real data from the \textit{MovieLens} study, which contains metadata on  44,879 movies released until July 2017. The data included a wide range of characteristics for each movie, such as cast, crew, budget, revenue, movie plot keywords among others. The goal was to predict the number of user reviews for a movie based on this available data. The histogram of numbers of users' reviews is given in Figure \ref{fig:User_rev_hist}. 

Due to the different types of data (numerical, text, JavaScript Object Notation format, etc.), significant effort was invested in feature preprocessing, resulting in a set of 276 features for each movie.

To model $d \sim n$ setup, we began  by randomly selecting a 5\% subset from the original large dataset ($n=2294$) and splitting it into training (1/3) and test (2/3) sets. We fitted first a log-linear Poisson regression for the saturated model with all 276 features. The analysis of results indicated the presence of overdispersion, so we continued with a NB regression model using \textit{glm.nb} function from the package \textit{MASS} \cite{MASS} with all 276 features. The \textit{glm.nb} also estimated the overdispersion parameter $\alpha$ using the method of moments, yielding $\hat{\alpha} = 2.2$ — a clear indication of overdispersion. The asymptotic test proposed in \cite{Agresti} for testing the null hypothesis $H_0: \frac{1}{\alpha} = 0$ (no overdispersion, i.e., the Poisson model) rejected the null hypothesis with a $p$-value $\ll$ 0.05.

To apply LASSO and SLOPE with $\gamma_j$'s from (\ref{eq:gamma}) for NB regression model we had also to adapt the existing FISTA algorithms available for Poisson regression to this case (see Appendix). Similar to the simulation study, the corresponding tuning parameters for all procedures were chosen by a 5-fold cross-validation.

We calculated (normalized) deviances $D^*(X\hat{\beta})=\frac{2}{n}KL({\bf Y},X\hat{\beta})$ for training and test sets, where from (\ref{eq:NB reg KL}),
$$
KL({\bf Y},X\hat{\beta}) =\sum_{i=1}^{n} \left(Y_i \ln\left(\frac{Y_i}{\exp(\hat{\beta}^t {\bf x}_i)}\right)-(Y_i+\alpha)\ln\left(\frac{Y_i+\alpha}{\exp(\hat{\beta}^t {\bf x}_i)+\alpha}\right)\right).
$$

The results are presented in Table \ref{tab:table_performance_sample}.
Similar to simulation studies, SLOPE implied the best prediction results in terms of deviance, both SLOPE and LASSO outperformed forward selection for both training  and test  sets.
On the other hand, 
SLOPE resulted again in the largest model with 97 features, while LASSO selected 55 features and the forward selection only 11. 

\begin{table}[h]
\begin{center}
\begin{tabular}{|l|c|c|c|}
  \hline
  Method & \# features & Training set deviance & Test set deviance\\
 \hline
 SLOPE & 97 & 1.273 & 1.409 \\
 LASSO & 55 & 1.299 & 1.418 \\
 Forward selection & 11 & 1.302 & 1.451 \\
 \hline
 \end{tabular}
\caption{Normalized deviances (5\% movies data subset).}
\label{tab:table_performance_sample}
\ignore{
 SLOPE   & 97  &   1095.8 & 2020.6\\
 LASSO &   55 & 1116.9 & 2034.12\\
 Forward Selection & 11 &  1119.745 & 2081.2 \\
 }
\end{center}
\end{table}

We then applied the three procedures to the entire dataset of 44,879 movies splitting them randomly into the training and test sets with respectively 75\% and 25\% of all movies. There was still clear evidence of overdispersion in the Poisson model ($\hat{\alpha} = 1.7$), so we used again the Negative Binomial regression model. The results, summarized in Table \ref{tab:table_performance_entire},  show that both SLOPE and LASSO performed similarly and selected almost identical models, differed by only two predictors. Both methods outperformed forward selection on both the training and test sets.

\begin{table}[h] 
\begin{center}
\begin{tabular}{|l|c|c|c|}
  \hline
  Method & \# features & Training set deviance & Test set deviance\\
  \hline
 SLOPE & 265 & 1.046 & 1.049 \\
 LASSO & 264 & 1.046 & 1.050 \\
 Forward selection & 11 & 1.183 & 1.165 \\
 \hline
\ignore{
 \hline
 SLOPE   & 265  &   36006.91 & 12031.63 \\
 LASSO &   264 & 35992.85 & 12047.68 \\
 Forward Selection & 11 &  40714.85 & 13359.97 \\
 }
\end{tabular}
\caption{Normalized deviances (the entire movies data set).}
\label{tab:table_performance_entire}
\end{center}
\end{table}

As expected, the goodness-of-fit improved with an increasing sample size, and the normalized deviances for both training and test sets decreased as the sample size grew. 
At the same time, compared to the results from the 5\% sample subset, the model sizes for both SLOPE and LASSO models significantly increased. This can be naturally explained by the reduction in the variances of the estimated coefficients as the sample size increases. As a result, even the weak effects of features that do not manifest and are not detected in a small sample become apparent and significant in a larger one. In contrast, forward selection identified the same subset of 11 highly significant features.


\noindent
\newline{\bf Acknowledgments.} The work was supported by the Israel Science Foundation (ISF), Grant ISF-1095/22. The authors 
are grateful to Amir Beck for his valuable remarks on FISTA.

\printbibliography

@article{SPRWL,
author = {Jinzhu, J. and Fang, X. and Lihu, X.},
title = {{Sparse Poisson regression with penalized weighted score function}},
volume = {13},
journal = {Electronic Journal of Statistics},
number = {2},
publisher = {Institute of Mathematical Statistics and Bernoulli Society},
pages = {2898 -- 2920},
year = {2019},
doi = {10.1214/19-EJS1580},
URL = {https://doi.org/10.1214/19-EJS1580}
}

@article{Figueiredo,
%     author = {Figueiredo, M. A. and Nowak, R. D.},
%     title = {Sparse estimation with strongly correlated variables using ordered weighted l1 regularization},
%     year = {2014},
%     DOI = "arXiv:1409.4005 [stat.ML]",
%     keywords = "owl"
% }

@article{10.1371/journal.pone.0209923,
    author = {Lehman, R. R. AND Archer, K. J.},
    journal = {PLOS ONE},
    % publisher = {Public Library of Science},
    title = {Penalized negative binomial models for modeling an overdispersed count outcome with a high-dimensional predictor space: Application predicting micronuclei frequency},
    year = {2019},
    volume = {14},
    number = {1},
    pages = {1-21}
}

@article{minimax_in_glm,
    author = {Abramovich, F. and Grinshtein, V.},
    journal = {IEEE Transactions on Information Theory},
    title = {Model selection and minimax estimation in generalized linear models},
    year = {2016},
    volume = {62},
    pages = {3721-3730},
    number = {6}
}

@article{LASSO_PAPER,
    author = {Tibshirani, R.},
    journal = {Journal of the Royal Statistical Society},
    series = {B (Methodological)},
    title = {Regression shrinkage and selection via the LASSO},
    year = {1996},
    volume = {58},
    pages = {267-288},
    number = {1}
}

@book{Agresti,
  title={Foundations of Linear and Generalized Linear Models},
  author={Agresti, A.},
  series={Wiley Series in Probability and Statistics},
  year={2015},
%   publisher={Wiley},
  pages = {248-250,269-270,307-310}
}

@article{LASSO_van_de_geer,
    author = {S. A. van de Geer},
    journal = {Annals of Statistics},
    title = {High-dimensional generalized linear models and the Lasso},
    year = {2008},
    volume = {36},
    pages = {614-645},
    number = {2}
}

@article{SLOPE_PAPER,
author = {Bellec, P. and Lecué, G. and Tsybakov, A.},
year = {2018},
pages = {3603 -- 3642},
title = {Slope meets Lasso: improved oracle bounds and optimality},
journal = {Annals of Statistics},
volume = {46},
number = {6B},
}

@article{FIRST_SLOPE,
author = {Bogdan, M. and Berg van den, E. and Sabatti, C. and Su, W. and Candès, E."},
journal = "Annals of Applied Statistics",
number = "3",
volume = "9",
pages = "1103--1140",
title = "SLOPE — adaptive variable selection via convex optimization",
year = "2015"
}

@book{McCullagh:1989,
  author = {McCullagh, P. and Nelder, J. A.},
  title = {Generalized Linear Models}, 
  edition = {Second},
  publisher = "Chapman \& Hall/CRC"}

@inproceedings{Larsson:2022,
author = {Larsson, J. and Bogdan, M. and Wallin, Jonas},
title = {The strong screening rule for SLOPE},
year = {2020},
isbn = {9781713829546},
publisher = {Curran Associates Inc.},
address = {Red Hook, NY, USA},
booktitle = {Proceedings of the 34th International Conference on Neural Information Processing Systems},
articleno = {1223},
numpages = {12},
location = {Vancouver, BC, Canada},
series = {NIPS '20}
}

@Article{Li:2022,
author = {Li, S. and Wei, H. and Lei, X.},
TITLE = {Heterogeneous overdispersed count data regressions via double-penalized estimations},
JOURNAL = {Mathematics},
VOLUME = {10},
YEAR = {2022},
NUMBER = {10},
ARTICLE-NUMBER = {1700},
}

@book{Giraud:2015,
  author = {Giraud, C. },
  title = {Introduction to High-Dimensional Statistics},
  year = 2015
}

@article{High_dimension_classification,
    author = {Abramovich, F. and Grinshtein, V.},
    year = {2019},
    title = {High-dimensional classification by sparse logistic regression},
    journal = {IEEE Transactions on Information Theory },
    volume = {65},
    number={5},
    pages = {3068-3079}
}

@article{Wang,
author = {Wang, Z. and Shuangge, M. and Michael Z. and Chirag P. and Ching-Yun W. and Prasad D.},
title ={Penalized count data regression with application to hospital stay after pediatric cardiac surgery},
journal = {Statistical Methods in Medical Research},
volume = {25},
number = {6},
pages = {2685-2703},
year = {2016},
}

@ARTICLE{Zou05regularizationand,
    author = {Zou, H. and Hastie, T.},
    title = {Regularization and variable selection via the Elastic Net},
    journal = {Journal of the Royal Statistical Society, Series B},
    year = {2005},
    volume = {67},
    pages = {301--320}
}

@Manual{SLOPE_package,
    title = {{SLOPE}: Sorted L1 Penalized Estimation},
    author = {Larsson, J. and Wallin, J. and Bogdan, M. and 
      van den Berg, van den,E. and Sabatti,C. and Candes, E. and
      Patterson, E. and Su, W.},
    year = {2021},
    note = {R package version 0.3.3},
    url = {https://CRAN.R-project.org/package=SLOPE}
  }

@Inbook{AIC_paper,
author="Akaike, H.",
title="Information theory and an extension of the maximum likelihood principle",
bookTitle= {Selected Papers of Hirotugu Akaike},
year="1998",
% publisher="Springer New York",
pages="199--213"
}

@article{RIC_paper,
    author = {D. P. Foster and E. I. George},
    journal = {Annals of Statistics},
    title = {The risk inflation criterion for multiple regression},
    year = {1994},
    volume = {22},
    pages = {1947-1975},
    number = {4}
}

@article{dim_of_model,
    author = {G. Schwarz},
    journal = {Annals of Statistics},
    title = {Estimating the dimension of a model},
    % publisher = {Institute of Mathematical Statistics},
    year = {1978},
    volume = {6},
    pages = {461–464},
    number = {2}
}

@article{var_sel_overview,
author = {Fan, J. and Lv, J.},
year = {2010},
pages = {101-148},
title = {A selective overview of variable selection in high dimensional feature space},
volume = {20},
journal = {Statistica Sinica}
}

@Book{MASS,
    title = {Modern Applied Statistics with S},
    author = {Venables, W. N. and Ripley, B. D.},
    publisher = {Springer},
    edition = {Fourth},
    year = {2002}
    }

@article{glmnet_paper,
author = {Tibshirani, R. and Hastie, T. and Friedman, J.},
year = {2010},
pages = {1-22},
title = {Regularized paths for generalized linear models Via coordinate descent},
volume = {33},
number={1},
journal = {Journal of Statistical Software},
}

@article{FISTA,
author = {Beck, A. and Teboulle, M.},
title = {A fast iterative shrinkage-thresholding algorithm for linear inverse problems},
year = {2009},
% publisher = {Society for Industrial and Applied Mathematics},
volume = {2},
number = {1},
journal = {SIAM Journal on Imaging Sciences},
pages = {183–202},
}

\section*{Appendix: FISTA for solving SLOPE in NB regression model}

FISTA (Fast Iterative Shrinkage-Thresholding Algorithm) proposed by \cite{FISTA} is a modified version of ISTA (Iterative Shrinkage-Thresholding Algorithm) that accelerates the convergence rate. Both are proximal descent algorithms for solving convex optimization problems of the form
\begin{equation} \label{eq:ISTA}
\min_{\mathbf{x}} \{ f(\mathbf{x})+g(\mathbf{x})\},
\end{equation}
where  $f$ is a smooth convex function and $g$ is a non-smooth convex function. Originally designed for solving LASSO, where  $g(\mathbf{x})=||\mathbf{x}||_1$, a general ISTA algorithm can be defined for any convex $g(\mathbf{x})$ as follows (see \cite{FISTA}).
Instead of minimizing (\ref{eq:ISTA}) directly, ISTA approximates the smooth $f$ by a quadratic approximation while keeping $g$ fixed and iteratively seeks
\begin{equation} 
\begin{split}
\mathbf{x}^{k+1} & = \arg\min_{\mathbf{x}}\left\{f(\mathbf{x}^k) + \nabla f(\mathbf{x}_k)^t(\mathbf{x} - \mathbf{x}_k) + \frac{1}{2t}||\mathbf{x} - \mathbf{x}_k||_2^2 + g(\mathbf{x})\right\} \\
& = \arg\min_{\mathbf{x}} \left\{\frac{1}{2t}  ||\mathbf{x}-(\mathbf{x}_k - t \nabla f(\mathbf{x}_k)||_2^2+g(\mathbf{x})\right\} \label{eq:proxy}
\end{split}
\end{equation}
for some $t$ (see \cite{FISTA}).

Define the proximal mapping $prox_g(\mathbf{x}) = \arg\min_{\mathbf{u}}\left\{\frac{1}{2}||\mathbf{u}-\mathbf{x}||_2^2+g(\mathbf{u})\right\}$. Then, the right hand side of (\ref{eq:proxy}) is $prox_{t g}(\mathbf{x}_k- t \nabla f(\mathbf{x}_k))$, and 
$\mathbf{x}^{k+1}$ is iteratively updated as
$$
\mathbf{x}^{k+1} = prox_{t g}(\mathbf{x}^k - t \nabla f(\mathbf{x}^k))
$$
  with the update step $t=1/L(f)$, where $L(f)$ is the Lipschitz constant of the gradient $\nabla f$. ISTA converges at the rate $O(k^{-1})$ \cite{FISTA}.

FISTA accelerates ISTA by applying the proximal mapping not at the previous point $\mathbf{x}_k$, but rather at the smartly chosen linear combination of $\mathbf{x}_k$ and $\mathbf{x}_{k-1}$. Namely,  
$$
\mathbf{w}^k = \mathbf{x}^k + \frac{\delta_{k} - 1}{\delta_{k+1}}(\mathbf{x}^k - \mathbf{x}^{k-1}),
$$
$$
\mathbf{x}^{k+1} = prox_{t g}(\mathbf{w}^k - t \nabla f(\mathbf{w}^k)),
$$
where $\delta_{k+1} = \frac{1+ \sqrt{1+ 4\delta_k^2}}{2}$ .
FISTA accelerates ISTA and improves the convergence rate to $O(k^{-2})$ (\cite{FISTA}).

We now apply the general FISTA algorithm to solving SLOPE in the NB regression model (\ref{eq:loglike_NB}),(\ref{eq:slope}) (recall that LASSO is a particular case of SLOPE with constant $\gamma$). In this context,  $f(\mathbf{\beta})=-l^{NB}(\mathbf{\beta})$ and $g(\mathbf{\beta})=\sum_{j=1}^d \gamma_j |\beta|_{(j)}$, where $\gamma_1 \geq \dotsc \geq \gamma_d \geq 0$.
By straightforward calculus, the gradient $\nabla(-l^{NB}) = \alpha \mathbf{X}^t \big(\frac{e^{\mathbf{\beta}^t \mathbf{x_i}} - y_i}{e^{\mathbf{\beta}^t \mathbf{x_i}} + \alpha}\big)_{i=1}^n
$ with the Lipschitz constant  $L(l^{NB})$ approximated by  
$L=\frac{(\alpha+\bar{\mathbf{y}})}{4\alpha} \cdot  \lambda_{max}\bigg( \mathbf{X}^t\mathbf{X} \bigg)$,
where $\lambda_{max}\bigg( \mathbf{X}^t\mathbf{X} \bigg)$ is the largest eigenvalue of $\mathbf{X}^t\mathbf{X}$.
The corresponding  proximal mapping can be obtained in closed form as
$$
prox_{t g}(\mathbf{x} - t \nabla f(\mathbf{x})) = \tau_{t \mathbf{\gamma}}(\mathbf{x} - t\nabla f(\mathbf{x})),
$$
where $\tau_{t\gamma}(\mathbf{u})_j = (|u|_{(j)} - t\gamma_j)_+ \cdot {\rm sign}(u_j),\;j=1,\ldots,d$ denotes the soft thresholding operator applied entry-wise to the  vector of ordered $|u|_j$'s with thresholds $t\gamma_j$'s.

The resulting FISTA for solving SLOPE in NB regression with a given sequence  $\gamma_1 \geq ... \geq \gamma_d \geq 0$ is given below: 
\begin{algorithm}
\SetKwInput{Kwinit}{Initialization}
\SetKwInput{KwGeneral}{General Step}
\SetKwInput{KwIn}{Inputs}
\SetAlgoLined
\KwIn{
     $f = \left \langle \alpha \mathbf{1}+\mathbf{y},\left( \ln \left(e^{\mathbf{\beta}^t \mathbf{x_i}} + \alpha\right) \right)_{i=1}^n \right\rangle - \bigg\langle \mathbf{y},\mathbf{X}\mathbf{\beta} \bigg\rangle $ \newline
     $g = \sum\limits_{j=1}^d{\gamma_j |\beta|_{(j)}}, \; \mathbf{\beta}^0 \in \mathbb{R}^d$ \newline
     $L = \frac{(\alpha+\bar{\mathbf{y}})}{4\alpha} \cdot  \lambda_{max} \bigg( \mathbf{X}^t\mathbf{X} \bigg)$
     }
\Kwinit{Set $\mathbf{w}^0 = \mathbf{\beta}^0, \quad \delta_0 = 1$}
\KwGeneral
 \BlankLine
 \Indp
 \For{$k = 0,1,\dotsc$ }
 {
  Sort $\mathbf{w}^k$'s in a descending order by their absolute values\; 
  Set \( \mathbf{\beta}^{k+1} =  
  \tau_{\frac{1}{L}\gamma}\left(\mathbf{w}^k - \frac{1}{L}\alpha \mathbf{X}^t \left(\frac{(e^{\mathbf{w}^k)^t} \mathbf{x}_i - y_i}{(e^{\mathbf{w}^k)^t \mathbf{x_i}} + \alpha}\right)_{i=1}^n\right)\) \; 
  Set $\delta_{k+1} = \frac{1+\sqrt{1+4\delta_k^2}}{2}$ \;
  Set $\mathbf{w}^{k+1} = \mathbf{\beta}^{k+1} + \frac{\delta_k-1}{\delta_{k+1}}(\mathbf{\beta}^{k+1} - \mathbf{\beta}^k)$ \; 
   }
\caption{FISTA for SLOPE in NB regression}
\label{FISTA_algorithm}
\end{algorithm}

\end{document}